\documentclass[twocolumn,showpacs,amsmath,amssymb,superscriptaddress]{revtex4-1}

\usepackage{graphicx}
\usepackage{dcolumn}
\usepackage{bm}
\usepackage[abs]{overpic}
\usepackage{here}
\usepackage{color}
\usepackage{url}
\usepackage{accents}
\def\up{\uparrow}
\def\down{\downarrow }

\newcommand{\wmax}{\ensuremath{{\omega_\mathrm{max}}}}

\begin{document}


\title{
Smooth self-energy in the exact-diagonalization-based dynamical mean-field theory:
Intermediate-representation filtering approach
}
\author{Yuki Nagai}
\affiliation{CCSE, Japan  Atomic Energy Agency, 178-4-4, Wakashiba, Kashiwa, Chiba, 277-0871, Japan}
\affiliation{
Mathematical Science Team, RIKEN Center for Advanced Intelligence Project (AIP), 1-4-1 Nihonbashi, Chuo-ku, Tokyo 103-0027, Japan
}

\author{Hiroshi Shinaoka}
\affiliation{Department of Physics, Saitama University, Saitama 338-8570, Japan}

\date{\today}
             
\begin{abstract}
We propose a method for estimating smooth real-frequency self-energy in the dynamical mean-field theory with the finite-temperature exact diagonalization (DMFT-ED).
One of the benefits of DMFT-ED calculations is that one can obtain real-frequency spectra without a numerical analytic continuation. 
However, these spectra are spiky and strongly depend on the way to discretize a continuous bath (e.g., the number of the bath sites).
The present scheme is based on a recently proposed compact representation of imaginary-time Green's functions,
the intermediate representation (IR).
The projection onto the IR basis acts as a \textit{noise} filter for the discretization errors in the self-energy.
This enables to extract the physically relevant part
from the \textit{noisy} self-energy.
We demonstrate the method for single-site DMFT calculations of the single-band Hubbard model.
We also show the results can be further improved by numerical analytic continuation.
\end{abstract}

\maketitle

\section{Introduction}
Strongly correlated electron systems have attracted much attention in the past 30 years, because of the discovery of
 interesting materials such as high temperature superconductors or heavy fermions. 
 These materials have rich phase diagrams, which have superconducting, magnetic, or Mott insulating phases. 
New theoretical schemes and techniques have been developed to understand many-body physics due to correlation between electrons in condensed matter physics. 

The dynamical mean-field theory (DMFT) is one of the most powerful tools to take strong correlations into account\cite{Georges}. 
The DMFT is used both in calculations with model Hamiltonians and in realistic electronic structure calculations\cite{Kotliar}.  
Within the DMFT, local dynamical correlations are fully preserved while the spatial correlations are neglected. 
Under this approximation, one has to solve an effective Anderson impurity model mapped from an original lattice model.  
The mapping onto the impurity model is enforced by a self-consistency condition which contains the information about the original lattice. 
Recently, it is known that the continuous-time quantum Monte Carlo method is one of the powerful ``exact'' impurity solvers since there is no discretization of the imaginary-time interval\cite{Gull}. 
One needs, however, to perform numerical analytical continuation to transform calculated imaginary-time quantity to real-frequency spectra, which is extremely sensitive to noise.
This is because the self-energy is computed in Matsubara frequency domain and must be continued to the real frequency axis.
QMC calculations also suffer from a negative sign problem at low temperature in solving cluster-type and multi-orbital impurity problems.

Exact diagonalization is another powerful ``exact'' and ``sign-problem-free solver" for the effective impurity model~\cite{Georges, Capone}. 
The ED can be combined with DMFT as the finite-temperature ($T$) DMFT-ED,
where the self-consistent procedure is implemented on the Matsubara-frequency axis.
In the finite-$T$ DMFT, 
the continuous bath of an impurity model is discretized with a finite number of bath sites.
Solving this approximate finite-size system by ED
provides direct access to quantities such as Green's functions and self-energies on the real-frequency axis without unstable numerical analytic continuation.
This is considered to be an advantage of the DMFT-ED.
The method has been used to investigate the real-frequency structure of the self-energy of the 2D Hubbard model at finite $T$~\cite{Sakai:2016jx} or effectively at $T = 0$ (i.e., sufficiently low $T$)~\cite{Civelli:2009gy,Kyung:2009fv}.

In DMFT-ED calculations, results of Matsubara/imaginary-time quantities converge quicker with a few number of bath sites per a correlated site/orbital~\cite{Capone}.
In contrast, real-frequency quantities have spiky structures, whose positions strongly depend on the number of the bath sites.
This strong bath-site dependence 
makes it difficult to interpret the physical meanings of their fine features.
The previous studies employed some artificial broadening (e.g. Lorentzian broadening)~\cite{Sakai:2016jx,Kyung:2009fv} or extrapolation to the real axis~\cite{Liebsch:2011br,Liebsch:in}  to obtain smooth results.

In this paper, we propose a method based on a physical ground for computing a {\it bath-discretization-insensitive} smooth self-energy on the real-frequency axis.
This is achieved by projecting the self-energy onto a physical compact basis and filtering out \textit{noise} due to the discretization of a continuous bath.
The present method is based on the recently shown fact that
the imaginary-time/Matsubara Green's functions are sparse at finite $T$:
the imaginary-time/frequency dependence of these Green's functions can be represented by a few dozen basis functions of the so-called intermediate representation (IR) basis~ \cite{Otsuki,ShinaokaPRB,ShinaokaPRB2,Chikano}.
In the previous study~\cite{Otsuki}, they used this representation together with sparse modeling (SpM) techniques in data science to extract 
noise-insensitive spectral functions from QMC data\cite{Otsuki}.
Using the IR, 
in the present study, 
we clarify the difference between quantities defined on the Matsubara-frequency and real-frequency axes in DMFT-ED calculations.
In particular, we use the IR-filtering approach 
to obtain the bath-number-insensitive spectrum in DMFT-ED calculations.
{To illustrate the procedure,}
we apply the present method to single-site DMFT calculations of the single-orbital Hubbard model.
We find that low-order coefficients of the self-energy in terms of the IR basis are converged with respect to the number of sites.
We demonstrate that the physically-relevant spectrum of the self-energy can be extracted from DMFT-ED calculations with  four bath sites.
The breaking of the causality in the reconstructed spectrum can be cured by imposing the non-negative condition on the spectrum using the SpM techniques.

\section{Sparseness of the self-energy}
\begin{figure}[t]
\begin{center}
     \begin{tabular}{p{ 0.9 \columnwidth}} 
      \resizebox{0.9  \columnwidth}{!}{\includegraphics{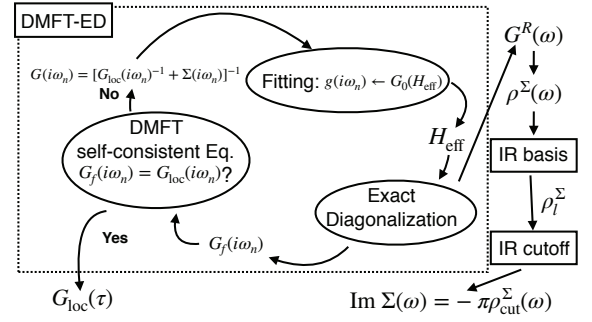}} 
    \end{tabular}
\end{center}
\caption{
\label{fig:DMFT}
Schematic figure of the calculation process to obtain the smooth self-energy $\Sigma(\omega)$ in the DMFT-ED calculation. 
}
\end{figure}

The DMFT method maps a lattice model onto an effective impurity model. 
Figure~\ref{fig:DMFT} illustrates the DMFT-ED self-consistent cycle for finite $T$.
The self-consistent equation reads
\begin{align}
G_{\rm loc}(i \omega_{n}) = G_{f}(i \omega_{n}), \label{eq:self}
\end{align}
where $G_{\rm loc}(i \omega_{n})$ and $G_{f}(i \omega_{n})$ are the local Matsubara temperature 
Green's functions of the original and effective impurity models, respectively. 
Here, we consider fermion Matsubara frequencies $\omega_{n} = \pi T (2n + 1)$ with $n = 0, \pm 1, \cdots, \pm n_{\rm max}$ ($n_{\rm max}$ is typically larger than $10^3$).
As revealed in the previous studies~\cite{Otsuki,ShinaokaPRB,Chikano}, however, $G_{\rm loc}(i \omega_{n})$ and $G_{f}(i \omega_{n})$ actually contain much fewer degrees of freedom than $n_\mathrm{max}$.  

The real-frequency self-energy is computed as
\begin{align}
\Sigma^R(\omega ) &\equiv {\cal G}_{0}^{R}(\omega)^{-1} - G^{R}(\omega)^{-1},\label{eq:real_freq_self}
\end{align}
$G^{R}(\omega )$ and ${\cal G}_{0}^{R}(\omega)$ are full and non-interacting retarded Green's functions for the effective impurity model, respectively. 
In the present study, we will clarify how much information is contained in this real-frequency self-energy.

In the previous study~\cite{Otsuki,ShinaokaPRB},
they considered an exact integral equation between the Matsubara Green's function $G(i\omega_n)$ and the spectral function $\rho(\omega)$ (i.e. Lehmann representation),
\begin{align}
G(i\omega_n) &= \int_{-\infty}^{\infty} d\omega K^\mathrm{F}(i\omega_n, \omega) \rho(\omega), \label{eq:gkr}
\end{align}
where $K^\mathrm{F}(\omega_n, \omega) \equiv 1/(i \omega_n - \omega)$ for fermions.

In the present study, we focus on a similar spectral representation for the self-energy~\cite{Luttinger}:
\begin{align}
\Sigma(i\omega_n) &= \Sigma_\mathrm{const}+\int_{-\infty}^{\infty} d\omega K^\mathrm{F}(i\omega_n, \omega) \rho^{\Sigma}(\omega), \label{eq:sigmar}
\end{align}
where $\rho^{\Sigma}(\omega) (\equiv -\pi^{-1}{\rm Im}(\Sigma^R(\omega)))$ is a spectral function and $\Sigma_\mathrm{const}$ is a frequency-independent term. 
Although $\rho^{\Sigma}(\omega)$ is a spiky bath-number-dependent function in the DMFT-ED framework,
$\Sigma(i\omega_n)$ computed by Eq.~(\ref{eq:sigmar}) is a smooth function.
This is because the model-independent filter matrix $K(i\omega_n, \omega)$ smears out fine features in $\rho^{\Sigma}(\omega)$.

This can be clearly seen by expanding these quantities in terms of ``intermediate representation'' (IR) basis \cite{ShinaokaPRB,Chikano} as 
\begin{align}
\Sigma(i\omega_n) - \Sigma_\mathrm{const} &= \sum_{l=0}^\infty  \Sigma_{l} U^\mathrm{F}_l(i \omega_n), \label{eq:sigmal}\\
\rho^{\Sigma}(\omega) &= \sum_{l=0}^\infty  \rho^{\Sigma}_{l} V^\mathrm{F}_l(\omega).\label{eq:rhol}
\end{align}
These basis functions are defined through the decomposition of the kernel as \cite{ShinaokaPRB,Chikano}
\begin{align}
	K^\mathrm{F}(i\omega_n, \omega) &= -\sum_{l=0}^{\infty} S^\mathrm{F}_l U^\mathrm{F}_l(i\omega_n) V^\mathrm{F}_l(\omega),\label{eq:kernel-exp}
\end{align}
The basis functions are orthonormalized in the intervals of $n\in [-\infty,+\infty]$ and $[-\wmax,\wmax]$, respectively ($\wmax$ is a cutoff frequency).
Note that the real part of the self-energy $\mathrm{Re}(\Sigma^\mathrm{Re}(\omega))$ can be computed from $\rho^{\Sigma}(\omega)$ by the Kramers-Kronig relation.

 \begin{figure}[t]
\begin{center}
     \begin{tabular}{p{1.0 \columnwidth}} 
      \resizebox{ 1.0 \columnwidth}{!}{\includegraphics{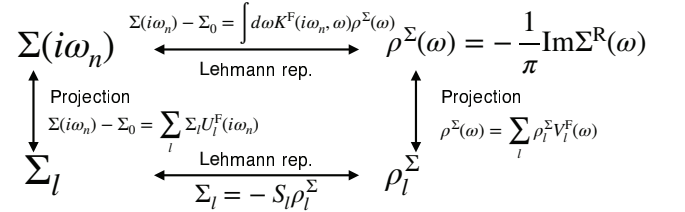}} 
    \end{tabular}
\end{center}
\caption{
\label{fig:relation}
Relation between different representations of the self-energy. 
}
\end{figure}

If $\wmax$ is large enough to cover the support of the spectrum,
$\rho^{\Sigma}_{l}$ and $\Sigma_l$ are related (see Fig.~\ref{fig:relation}) as 
\begin{align}
\Sigma_{l} &= -S^\mathrm{F}_{l} \rho^{\Sigma}_{l}.\label{eq:gsrs}
\end{align}
Because the singular values $S^\mathrm{F}_{l}$ $(l  = 0,1,2,\cdots)$ decay exponentially,
in numerical calculations with double precision floating point numbers, any Green's functions and self-energies can be represented by only few $U_l(i\omega_n)$ corresponding to large singular values (i.e. $S_l/S_0 > 10^{-8}$). 
In the present study, we use $\beta = 1/T = 20$ and $\wmax = 5$. 
This leaves only 25 ($=l_\mathrm{max}$) singular values.

The above consideration indicates that the Green's functions in Eq.~(\ref{eq:self}), and hence the real-frequency self-energy, have few independent variables no more than $l_{\rm max}$,
which does not depend on details of the model. 
As clarified below, this explains why the Matsubara 
Green's function converges to that obtained by QMC calculation with a small number of bath sites in DMFT-ED calculations~\cite{Capone}.

On the basis of the concept of the IR, we may obtain a converged solution
{\it only if the variables of the effective Hamiltonian is more than the number of singular values $S_{l}$ required for a desired accuracy.}
The effective impurity Hamiltonian reads
 \begin{align}
 H_{\rm AIM} = \sum_{l\sigma} \epsilon_{l\sigma} c_{l \sigma}^{\dagger} c_{l \sigma} + \sum_{l \sigma} V_{l \sigma} (f^{\dagger}_{\sigma} c_{l \sigma}+ c_{l \sigma}^{\dagger} f_{\sigma})+ H_{\rm atomic},\label{eq:effective_model}
 \end{align}
 where $H_{\rm atomic}$ the on-site atomic part of the original Hamiltonian. For example, the Hubbard model has $H_{\rm atomic} = -\mu \sum_{\sigma} f_{\sigma}^{\dagger} f_{\sigma} + U  f_{\up}^{\dagger} f_{\up} f_{\down}^{\dagger} f_{\down}$. 
Here, $f^{\dagger}_{\sigma}$ and $c_{l \sigma}$ are creation operators for fermions in with spin $\sigma$ associated with the impurity site and with the state $l$ of the effective bath, respectively. 
Thus, the effective impurity Hamiltonian involves $N_s$ independent parameters ($\epsilon_{l\sigma}$ and $V_{l \sigma}$).

\begin{figure}[t]
\begin{center}
     \begin{tabular}{p{ 0.9 \columnwidth}} 
      \resizebox{0.9  \columnwidth}{!}{\includegraphics{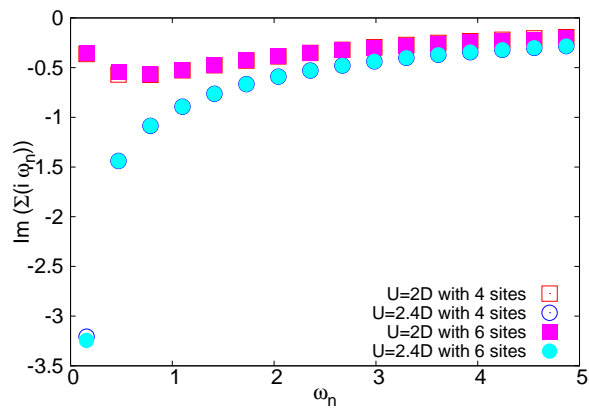}} 
    \end{tabular}
\end{center}
\caption{
\label{fig:selfmatsubara}
(Color online) Imaginary part of the self-energy on the Matsubara frequency axis ${\rm Im} \: \Sigma(i \omega_n)$ with $U=2D$ and $U=2.4D$ after the 200 loops of the DMFT-ED calculation with 4 site baths. 
We consider $\beta = 20$, $\omega_{\rm max} = 5$, $n_{\rm max} = 1024$ in the half-filled Hubbard model with the semicircular density of states with a half bandwidth $D$.  
}
\end{figure}

 \begin{figure}[t]
\begin{center}
     \begin{tabular}{p{1 \columnwidth}} 
      (a)\resizebox{0.9  \columnwidth}{!}{\includegraphics{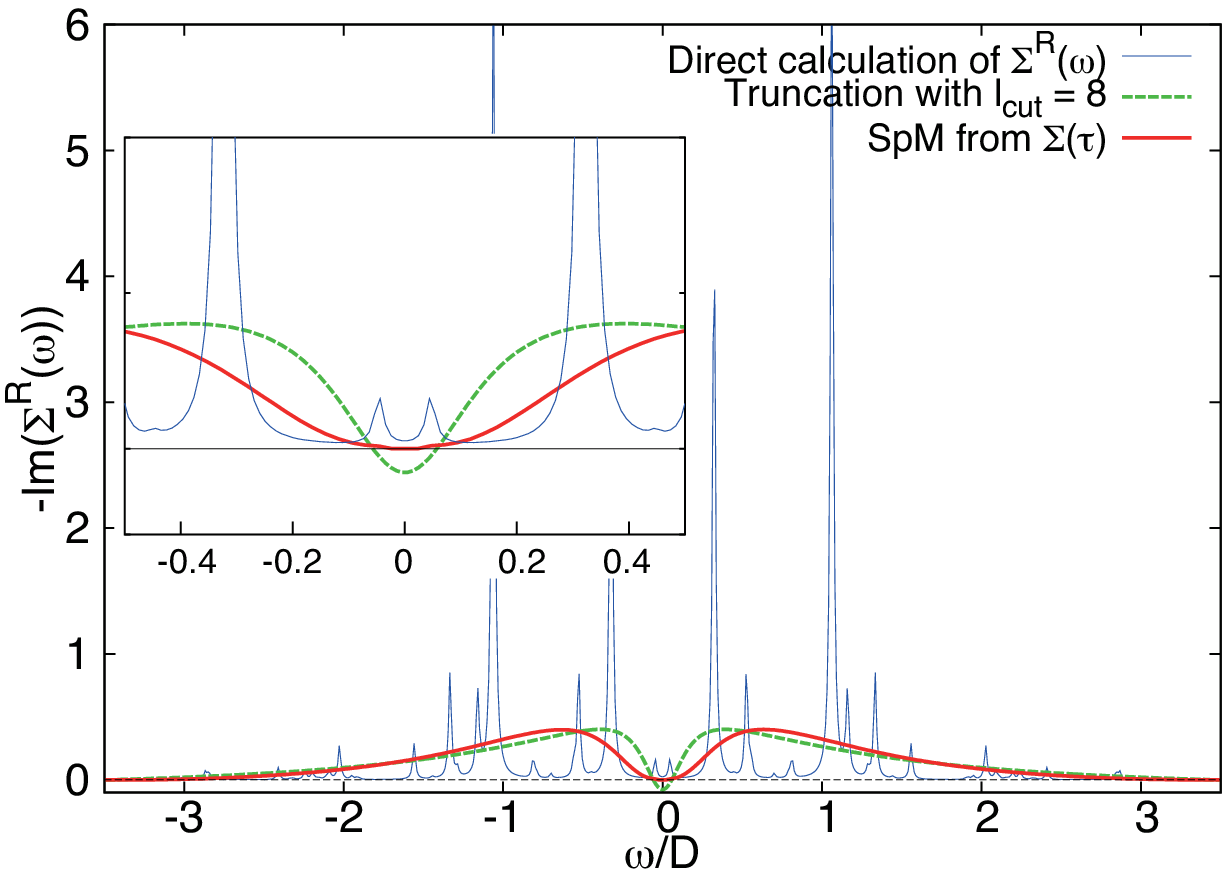}} 
      \\ 
      (b)\resizebox{0.9 \columnwidth}{!}{\includegraphics{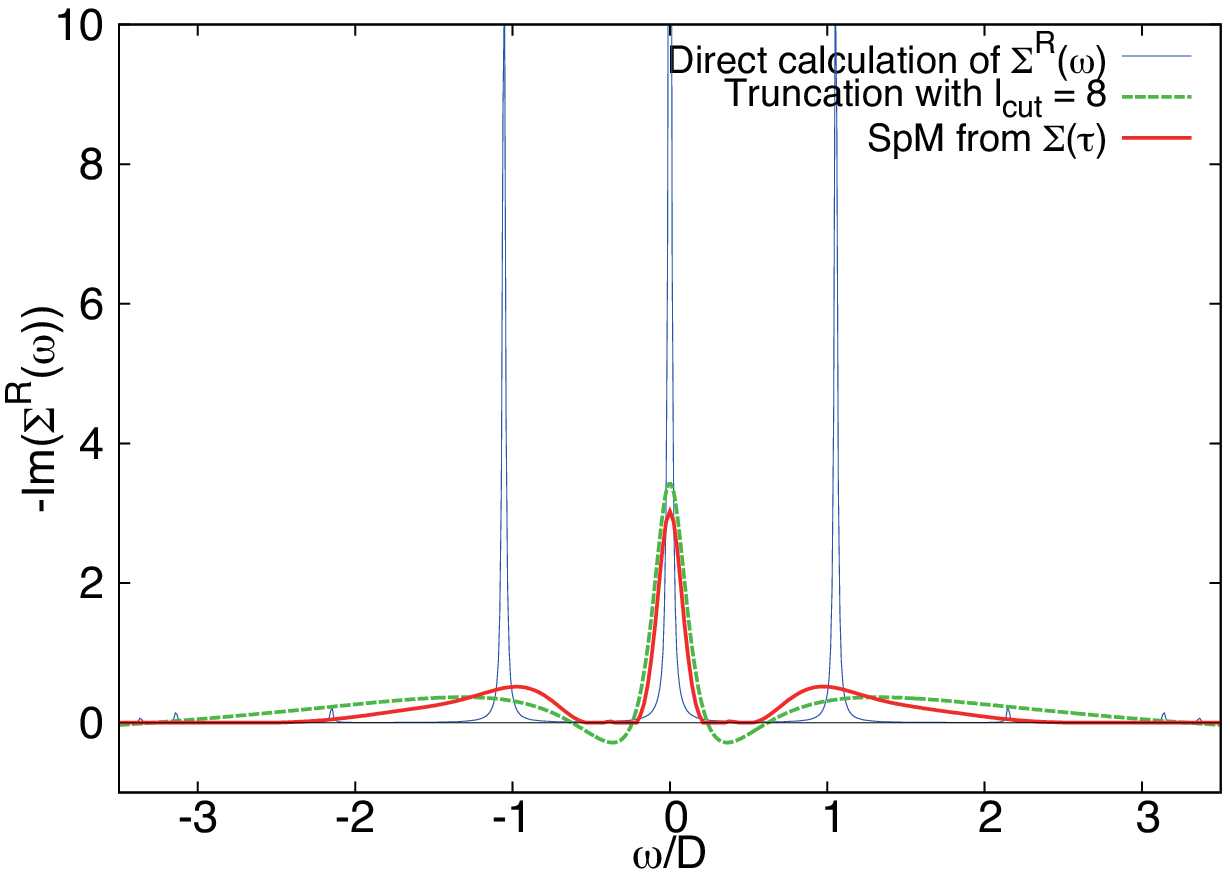}} 
    \end{tabular}
\end{center}
\caption{
\label{fig:IRself}
(Color online) Imaginary part of the self-energy ${\rm Im} \Sigma^R(\omega)$ calculated by the direct calculation in the DMFT-ED, the truncation with $l_{\rm cut}=8$, SpM analytic continuation for (a) $U=2D$ and (b) $U = 2.4D$. 
We consider 4 bath sites. 
Other parameters are the same as in Fig.~\ref{fig:selfmatsubara}.
}
\end{figure}
\begin{figure}[t]
\begin{center}
     \begin{tabular}{p{1.0 \columnwidth}} 
      \resizebox{ 1.0 \columnwidth}{!}{\includegraphics{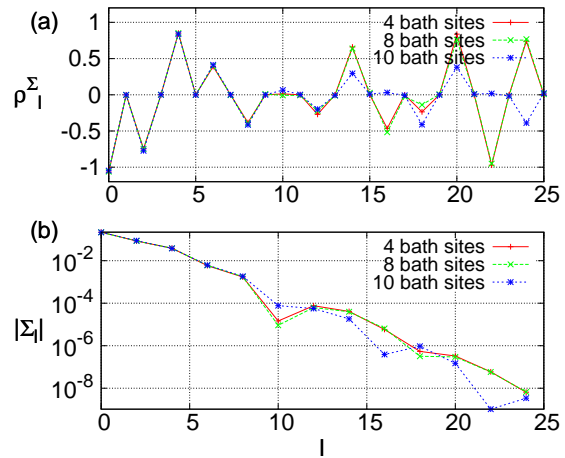}} 
    \end{tabular}
\end{center}
\caption{
\label{fig:sigmal}
(Color online) Bath-site-number dependence of the coefficients $ \rho^{\Sigma}_{l}$ and $|\Sigma_{l}|$ of IR basis after the 200 loops of the DMFT-ED calculation. 
Other parameters are the same as in Fig.~\ref{fig:selfmatsubara}.
}
\end{figure}

\section{\textit{Extracting physically-relevant spectrum from the self-energy}}
Now we propose a method for reconstructing a smooth self-energy on the real-frequency axis by extracting physically-relevant available information in the self-energy computed with a finite $N_s$.
In particular, 
we consider a  half-filled single-band Hubbard model with the semicircular density of states $\rho_0(\omega) = [2/(\pi D)\sqrt{1-(\omega/D)^2}]$ as the original lattice model. We set the half bandwidth $D$ as the energy unit. 
In particular, we focus on metallic and insulating solutions obtained for $U=2D$ and $2.4D$, respectively.
For the present model, $\Sigma_\mathrm{const} = 0$ due to particle-hole symmetry.
We describe the technical details of the bath fitting in Appdendix~\ref{appendix:fitting}.

Figure~\ref{fig:selfmatsubara} and \ref{fig:IRself} show the computed $\mathrm{Im}(\Sigma(i\omega_n))$ and  $\rho^{\Sigma}(\omega)$ ($=-\frac{1}{\pi} {\rm Im}\Sigma^R(\omega)$), respectively.
As clearly seen, $\mathrm{Im}(\Sigma(i\omega_n))$ is a smooth function, being almost bath-site-number independent.
In contrast, $\rho^{\Sigma}(\omega)$ computed by Eq.~(\ref{eq:real_freq_self}) has spiky structures, which strongly depend on the number of the bath sites (not shown).

To obtain a deeper insight into this contrasting behavior,
we compute their expansion coefficients in the IR using Eqs.~(\ref{eq:sigmal}) and (\ref{eq:rhol}).
Figure~\ref{fig:sigmal} compares $\Sigma_{l}$ and $\rho^{\Sigma}_l$ computed with $N_s = 4, 8, 10$ for $U=2D$.
As clearly seen in Fig.~\ref{fig:sigmal}(a), $\rho^{\Sigma}_l$ does not vanish as $l$ is increased.
Furthermore,  the large-$l$ components strongly depends on $N_s$.
On the other hand, $\Sigma_l$ decay exponentially with respect to $l$.
While the coefficients $\Sigma_{l}$ for small $l$ ($\le 8 $) are converged only with four bath sites,
those for larger $l$ fluctuate and do not converge systematically with respect to $N_s$.
These small discretization errors in $\Sigma_l$ are amplified and propagate to $\rho^{\Sigma}_l$ through Eq.~(\ref{eq:gsrs}), leading to the large fluctuations of $\rho^{\Sigma}_l$ at large $l$.
This strongly indicates that the bath-site-dependent (spiky) component of the spectral function originates from tiny discretization errors in the Green's function.
Thus, they {\it are not} physically relevant in the finite-temperature DMFT calculation based on the Matsubara Green's function.

We now extract this physically relevant information in the imaginary-time self-energy to estimate a smooth self-energy on the real-frequency axis.
By truncating the expansion at $l = l_{\rm cut}$ (IR filtering), we obtain smooth self-energies on real frequency axis as shown in Fig.~{\ref{fig:IRself}}. 
With the use of these self-energies, we can understand difference between systems with $U=2D$ and $U=2.4D$. 
Note that the imaginary part of the self-energy is the inverse of the quasiparticle lifetime. 
In the case of $U=2D$, the density of states should have a peak at the zero energy since the imaginary part of the self-energy at zero energy is zero, which suggests that this system is metallic. 
In the case of $U=2.4D$, there is no intensity of the density of states at the zero energy since there is a strong quasiparticle dumping at the zero energy, which suggests that this system is in the Mott insulating phase. 
We note that the self-energy on the real-frequency axis is also useful for the non-hermitian topological theory in finite-temperature strongly-correlated systems\cite{Shen,Kozii,YNp}. 

As shown in Fig.~\ref{fig:IRself}(b), the self-energy with the truncation $l_{\rm cut} = 8$ has the non-causal region ($-{\rm Im}\Sigma^R(\omega) < 0$) around $|\omega|=0.3D$, which may be ascribed to Gibbs oscillations due to the truncation.
This can be fixed by using the SpM analytic continuation~\cite{Otsuki}.
This approach 
allows us to obtain a solution which satisfies non-negativity $\rho^{\Sigma}(\omega) > 0$ for the self-energy with the correct sum rule.
We shows the results obtained by analytic continuation in Fig.~\ref{fig:IRself}.
Not only causality is restored but also the correct Fermi-liquid-like low-$\omega$ behavior is recovered for $U=2D$.

\section{Discussions}
With the use of the IR basis,
we can discuss whether the number of the bath sites is enough or not. 
In cluster or multi-orbital DMFT-ED calculations, the number of the bath sites is usually limited to a small number. 
In these cases, one can discuss a validity by comparing two systems with different number of the bath sites. 

At finite temperature, the DMFT and imaginary-time Green function-based methods have maximally $l_{\rm cut}$ information, since 
the kernel $K(i \omega_{n},\omega)$ is a model-independent function. 
An energy resolution of a spectral function is determined by the temperature and spectral width. 
Even if the number of the bath sites is extremely large, the resolution of the Matsubara Green's function in Eq.~(\ref{eq:self}) is limited by the minimum singular value $s_{l} > \delta$.
In terms of the data science, there might be an over-fitting problem when the number of the bath site is much larger than the number of the effective singular values, since one needs to fit the sparse Matsubara Green's function to obtain the coefficients of the effective impurity Hamiltonian. 

In the ED impurity solver,
the exponential growth of the Hilbert space limits 
the total number of correlated sites/orbital and bath sites to a small number ($<16$).
Recently, extensive efforts have been made to treat more correlated (and thus more bath sites)
by somehow truncating the Hilbert space~\cite{Zgid11, Zgid12, Go15, Go17}.
Another interesting approach is based on imaginary-time evolution of matrix product states~\cite{Wolf:2015dr}.
In these methods,
the discretization of a continuous bath is formulated in imaginary time although they are aimed at zero-$T$ calculations.
Thus, they suffer from discretization errors in computing quantities on the real-frequency axis to some extent.
The present scheme applies to these methods.

\section{Summary}
In this work, we proposed the IR-filtering approach for computing the physically relevant self-energies in DMFT-ED calculations formulated in imaginary time.
The present scheme is based on the IR basis which acts as a filter for the discretization errors of a continuous bath.
We demonstrated the procedure for the single-site DMFT-ED calculations of the single-band Hubbard model.
We showed that the result can be further improved by using the SpM techniques of analytic continuation.
This approach can be applied to multi-orbital or cluster DMFT calculations.

\section*{Acknowledgments}
Y. N. would like to acknowledge S. Yamada for helpful discussions and comments about the exact diagonalization technique.
H. S. thanks S. Sakai, J. Otsuki, K. Yoshimi, N. Chikano for fruitful discussions.
We use ``irbasis 0.1.7'', a Python library for the intermediate representation (IR) basis functions\cite{ChikanoRev}. 
The irbasis is called in the Julia language 0.6.2.
The calculations were partially performed by the supercomputing systems SGI ICE X at the Japan Atomic Energy Agency. Y. N. was partially supported by 
JSPS KAKENHI Grant Number 18K11345 and 18K03552, the ``Topological Materials Science'' (No. JP18H04228) KAKENHI on Innovative Areas from JSPS of Japan. 
HS was supported by JSPS KAKENHI Grants No. 16H01064 (J-Physics), 18H04301 (J-Physics), 16K17735, 18H01158.

\appendix

\section{Details of the finite-temperature dynamical mean field theory with the exact diagonalization solver}
The DMFT method maps a lattice model onto an effective impurity model written as 
 \begin{align}
 H_{\rm AIM} = \sum_{l\sigma} \epsilon_{l\sigma} c_{l \sigma}^{\dagger} c_{l \sigma} + \sum_{l \sigma} V_{l \sigma} (f^{\dagger}_{\sigma} c_{l \sigma}+ c_{l \sigma}^{\dagger} f_{\sigma})+ H_{\rm atomic},
 \end{align}
 where $H_{\rm atomic}$ the on-site atomic part of the original Hamiltonian. For example, the Hubbard model has $H_{\rm atomic} = -\mu \sum_{\sigma} f_{\sigma}^{\dagger} f_{\sigma} + U  f_{\up}^{\dagger} f_{\up} f_{\down}^{\dagger} f_{\down}$. 
Here, $f^{\dagger}_{\sigma}$ and $c_{l \sigma}$ are creation operators for fermions in with spin $\sigma$ associated with the impurity site and with the state $l$ of the effective bath, respectively. 

The self-consistent equation in the DMFT framework is written as 
\begin{align}
G_{\rm loc}(i \omega_{n}) = G_{f}(i \omega_{n}),
\end{align}
where $G_{\rm loc}(i \omega_{n})$ is the local Matsubara Green's function of the original Hamiltonian defined as 
\begin{align}
G_{\rm loc}(i \omega_{n}) &= \int d\epsilon D(\epsilon) [i \omega_{n} + \mu - \epsilon - \Sigma(i \omega_{n})]^{-1}. 
\end{align}
The momentum-independent self-energy $\Sigma(i \omega_{n})$ is calculated by the Dyson equation for the effective impurity model: 
\begin{align}
\Sigma(i \omega_{n}) &\equiv {\cal G}_{0}^{-1}(i \omega_{n}) - G_{f}^{-1}(i \omega_{n}).
\end{align}
Here, ${\cal G}_{0}(i \omega_{n})$ is a non-interacting part of the Green's function for the effective impurity model calculated as 
\begin{align}
 {\cal G}_{0}^{-1}(i \omega_{n}) &= i \omega_{n} +\mu - \sum_{l}^{N_{s}} \frac{|V_{l}|^{2}}{i \omega_{n} - \epsilon_{l}}. 
\end{align}
The impurity Green's function is calculated as 
\begin{align}
G_{f,\sigma}(i \omega_{n}) &= -\frac{1}{Z} \sum_{k} e^{- \beta E_{k}} G_{\sigma}^{(k)}(i\omega_{n}),
\end{align}
where 
\begin{align}
&G_{\sigma}^{(k)}(i\omega_{n}) = \langle k | f_{\sigma} [i \omega_{n} - ( H_{\rm AIM} - E_{k})]^{-1} f_{\sigma}^{ \dagger}| k \rangle \nonumber \\
& +\langle k | f_{\sigma}^{ \dagger} [i \omega_{n} + ( H_{\rm AIM} - E_{k})]^{-1} f_{\sigma}| k \rangle. 
\end{align}
Here, $Z = \sum_{k} e^{- \beta E_{k}}$ is a partition function and $E_{k}$ and $|k \rangle$ is an $k$-th eigenstate and eigenvalue of $H_{\rm AIM}$, respectively. 
The local density of states is expressed as
\begin{align}
\rho_{\sigma}(\omega) = \frac{1}{Z} \sum_{k} e^{- \beta E_{k}} (1/\pi) {\rm Im} \: G_{\sigma}^{(k)}(i\omega_{n} \rightarrow \omega + i \eta). \label{eq:rho}
\end{align}

\section{Bath fitting scheme}~\label{appendix:fitting}
At each DMFT step, we  fit the parameters $\epsilon_{l}$ and $V_{l}$ to reproduce ${\cal G}_{0}(i \omega_{n})$, 
and compute the impurity Green's function $G_{f,\sigma}(i \omega_{n})$ by exact diagonalization.
Then, we compute the non-interacting Green's function ${\cal G}_{0}(i \omega_{n}) = [G_{f,\sigma}(i \omega_{n})^{-1} + \Sigma(i \omega_{n})]^{-1}$ for the next iteration.
In the fitting, we minimize the function defined as 
\begin{align}
f(\{ \epsilon_{l}\}, \{ V_{l} \}) &= \sum_n \Bigl{|} {\cal G}_{0}(i \omega_{n}) - \left[i \omega_n + \mu - \sum_l \frac{V_l^2}{i \omega_n - \epsilon_l} \right]^{-1} \Bigl{|}, 
\end{align}
with the use of the one of the quasi-Newton algorithms called Broyden-Fletcher-Goldfarb-Shanno (BFGS) method in Optim.jl package written in Julia language.
We used the lowest 1024 positive Matsubara frequencies in the fitting.

\section{Green's function obtained by the exact diagonalization technique at finite temperature}
We describe how to calculate Green's function as follows. 
To calculate the  impurity Green's function, we obtain the eigenvalues and eigenvectors of $H_{\rm AIM}$. 
In the Hubbard model, the Hamiltonian matrix $H_{\rm AIM}$ is block diagonal since the spin-number is conserved in each block matrices. 
Thus, the Green's function is expressed as 
\begin{align}
G_{f,\sigma}(i \omega_{n}) &= -\frac{1}{Z} \sum_{n_{\up}=0}^{N_{s}+1} \sum_{n_{\down} = 0}^{N_{s}+1} \sum_{l} e^{- \beta E_{n_{\up},n_{\down} }^{l}} G_{\sigma}^{(n_{\up},n_{\down},l)}(i\omega_{n}),
\end{align}
where 
\begin{align}
&G_{\up}^{(n_{\up},n_{\down},l)}(i\omega_{n})= 
\langle (n_{\up},n_{\down},l) |
 f_{\up}  \nonumber \\
&\times  [i \omega_{n} - ( H_{\rm AIM}^{n_{\up}+1,n_{\down}} - E_{n_{\up},n_{\down} }^{l})]^{-1}
  f_{\up}^{ \dagger}| (n_{\up},n_{\down},l) \rangle \nonumber \\
& +\langle (n_{\up},n_{\down},l) | f_{\up}^{ \dagger} \nonumber \\
&\times [i \omega_{n} + ( H_{\rm AIM}^{n_{\up}-1,n_{\down}} - E_{n_{\up},n_{\down} }^{l})]^{-1} f_{\up}| (n_{\up},n_{\down},l) \rangle. 
\end{align}
\begin{align}
&G_{\down}^{(n_{\up},n_{\down},l)}(i\omega_{n})= 
\langle (n_{\up},n_{\down},l) |
 f_{\down}  \nonumber \\
 &\times [i \omega_{n} - ( H_{\rm AIM}^{n_{\up},n_{\down}+1} - E_{n_{\up},n_{\down} }^{l})]^{-1}
  f_{\down}^{ \dagger}| (n_{\up},n_{\down},l) \rangle \nonumber \\
& +\langle (n_{\up},n_{\down},l) | f_{\down}^{ \dagger} \nonumber \\
&\times  [i \omega_{n} + ( H_{\rm AIM}^{n_{\up},n_{\down}-1} - E_{n_{\up},n_{\down} }^{l})]^{-1} f_{\down}| (n_{\up},n_{\down},l) \rangle. 
\end{align}
Here, $N_{s}$ is the number of bath sites, and $n_{\up}$ and $n_{\down}$ are numbers of up and down spins, respectively.
$H_{\rm AIM}^{n_{\up},n_{\down}}$ is the Hamiltonian matrix with fixed $n_{\up}$ and $n_{\down}$.  $E_{n_{\up},n_{\down} }^{l}$ and 
$| (n_{\up},n_{\down},l) \rangle$ are the $l$-th eigenvalues and eigenvectors associated with $H_{\rm AIM}^{n_{\up},n_{\down}}$. 
In this paper, we use eigs in the Julia language 0.6.2, which computes eigenvalues and eigenvectors using implicitly restarted Lanczos iterations for real symmetric matrix. 
The diagonal element of the inverse of the matrix $\langle (n_{\up},n_{\down},l) |
 f_{\up} [i \omega_{n} - ( H_{\rm AIM}^{n_{\up}+1,n_{\down}} - E_{n_{\up},n_{\down} }^{l})]^{-1}
  f_{\up}^{ \dagger}| (n_{\up},n_{\down},l) \rangle$ is calculated by the Lanczos method\cite{Georges}. 
In all calculations in main text, we consider 4 eigenvalues and eigenvectors in each block diagonal Hamiltonian $H_{\rm AIM}^{n_{\up},n_{\down}}$.

\clearpage
\newpage
\widetext
\onecolumngrid

\end{document}